\documentclass[12pt]{article}
\usepackage[a4paper, total={6.5in, 9in}]{geometry}
\usepackage{natbib}
\usepackage{amsmath}
\usepackage{amssymb}
\usepackage{amsthm}
\usepackage{graphicx}
\usepackage{setspace}
\usepackage{float}
\usepackage{enumitem}
\usepackage{xcolor}
\usepackage{placeins}
\usepackage{authblk}

\newtheorem{assumption}{Assumption}
\newtheorem{remark}{Remark}

\title{Counterfactual Density Effects and the German East--West Income Gap}

\author[1,*]{Georg Keilbar}
\author[1]{Sonja Greven}
\affil[1]{Chair of Statistics, Humboldt-Universit\"at zu Berlin, Germany}
\affil[*]{\textit{Corresponding author,} \texttt{georg.keilbar@hu-berlin.de}}

\date{\today}

\begin{document}
\doublespacing

\maketitle

\begin{abstract}
We propose a novel framework for conducting causal inference based on counterfactual densities. While the current paradigm of causal inference is mostly focused on estimating average treatment effects (ATEs), which restricts the analysis to the first moment of the outcome variable, our density-based approach is able to detect causal effects based on general distributional characteristics. Following the Oaxaca-Blinder decomposition approach, we consider two types of counterfactual density effects that together explain observed discrepancies between the densities of the treated and control group. First, the distribution effect is the counterfactual effect of changing the conditional density of the control group to that of the treatment group, while keeping the covariates fixed at the treatment group distribution. Second, the covariate effect represents the effect of a hypothetical change in the covariate distribution. Both effects have a causal interpretation under the classical unconfoundedness and overlap assumptions. Methodologically, our approach is based on analyzing the conditional densities as elements of a Bayes Hilbert space, which preserves the non-negativity and integration-to-one constraints. We specify a flexible functional additive regression model estimating the conditional densities. We apply our method to analyze the German East--West income gap, i.e., the observed differences in wages between East Germans and West Germans. While most of the existing studies focus on the average differences and neglect other distributional characteristics, our density-based approach is suited to detect all nuances of the counterfactual distributions, including differences in probability masses at zero.
\end{abstract}

\vspace{6pt}

\section{Introduction}\label{sec:introduction}

The most prevalent approaches in causal inference are based on the study of mean-based quantities such as the average treatment effect (ATE) and the average treatment effect on the treated (ATT). While it is true that ATEs are easy to interpret and can be estimated reliably in many situations, they are nonetheless restricted to identifying location-shifts in the data while ignoring effects that go beyond the first moment of the distribution. Recently, there has been growing interest in more nuanced approaches based on quantiles or other distributional characteristics \citep{chernozhukov2005iv, firpo2007efficient, chernozhukov2013inference}.

In this paper, we advocate for a density-focused approach for causal inference. Similar to quantiles, densities reflect the entire distribution of the variable of interest. However, densities offer some important advantages. The computational burden is reduced, as estimating conditional densities involves a single estimate, whereas distributional and quantile regressions must be run separately for different values of the threshold index and quantile level, respectively.
There is also no monotonicity issue in the estimation unlike in quantile-based approaches. Consequently, rearrangement methods to avoid quantile crossing, such as those proposed in \cite{chernozhukov2010quantile}, are not required.
Additionally, it can be argued that densities are better suited to display and intuitively understand the shape of the data compared to distribution and quantile functions, e.g., in the presence of bimodalities or shifts in the probability mass.
Finally, methods based on quantile regression usually require continuous distributions, while our density-based approach is also applicable in the case of discrete and mixed-type outcome variables. These problems are highly relevant in the case of income distributions considered in our empirical application, which exhibit both bimodalities and probability masses at zero.  

Our definition of counterfactual densities and the resulting density effects build on the decomposition literature, originating with \cite{blinder1973wage} and \cite{oaxaca1973male}. Specifically, we define counterfactual densities based on the conditional density of group A, evaluated as if it were exposed to the covariate distribution of another group, B. The original Oaxaca--Blinder decomposition was introduced for the purpose of explaining differences in means of two groups. A special case of the decomposition focusing on proportions and categorical covariates had already been proposed by \citet{kitagawa1955components}. For approaches beyond the mean, similar decompositions based on counterfactual quantities have been studied for quantiles and other distributional characteristics \citep{dinardo1996labor, firpo2018decomposing, chernozhukov2013inference}. We refer to \citet{fortin2011decomposition} for a comprehensive overview of decomposition methods in economics. These approaches are based on an \emph{additive} Oaxaca--Blinder type decomposition. As a crucial difference, we instead focus on a \emph{multiplicative} decomposition. We consider two types of counterfactual density effects that admit a causal interpretation under the standard unconfoundedness and overlap assumptions. First, the distribution effect captures the counterfactual impact of changing the conditional density from that of the control group to that of the treatment group, while holding the covariate distribution fixed at the treatment group level. Second, the covariate effect represents the effect of a hypothetical change in the covariate distribution, with the conditional density held fixed at the control group.

The estimation of counterfactual densities depends critically on accurate estimates of conditional densities. However, many existing approaches have important limitations. On the one hand, fully nonparametric estimators suffer from the curse of dimensionality, making them impractical even in moderately dimensional settings. Examples include the log-spline approach of \citet{stone1991asymptotics, stone1994use} and local polynomial-based estimators \citep{fan1996estimation, cattaneo2024boundary}. On the other hand, fully parametric models rely on strong distributional assumptions, which can lead to model misspecification when these assumptions are violated. In this paper, we instead rely on the Bayes Hilbert space approach for conditional density estimation of \cite{maier2024conditional}. The authors propose a flexible structured additive regression model that obeys the logic of densities, i.e., the estimated densities are non-negative and integrate to one. Estimation relies on the use of basis functions (e.g., splines) and is based on a Poisson approximation to the Bayes Hilbert space likelihood.

Our counterfactual density methodology is motivated by the analysis of the East--West income gap in Germany. This empirical application illustrates the advantages of density-based approaches for several reasons. First, some of the estimated densities exhibit heavy skewness and bimodality, features that are easily detectable in density plots but may be difficult to identify using estimated quantile or distribution functions. Second, the income variable is zero-inflated, with the fraction of zeros varying with the covariates, which poses challenges for quantile-based methods. This is particularly relevant since a focus on the positive part of the income distribution would avoid this problem only at the expense of losing information about the unemployed. Empirically, we find that the East--West gap has narrowed over the past 30 years after reunification; however, notable differences still persist. Our results suggest that these differences are largely driven by the conditional distribution rather than by differences in the composition of covariates. I.e., differences in the covariate distributions explain only a small part of the observed differences in the income distributions. Finally, we find that these differences are much more pronounced when focusing on the male subpopulation. We therefore conclude that the East--West income gap in Germany is, to a large extent, a male-specific issue.

Methods for counterfactual distributions based on quantile and distributional regression instead of density regression have been discussed before. \cite{chernozhukov2013inference} is one of the most closely related papers. They focus on distributional (i.e., for the cumulative distribution function) and quantile effects, but consider a similar Oaxaca--Blinder decomposition of effects. Instead of estimating the conditional density, their approach relies on quantile and distribution function regression. For both approaches, they rely on known basis functions of covariates. Similarly, \cite{machado2005counterfactual} consider an Oaxaca--Blinder decomposition based on estimating conditional quantiles. In particular, they estimate counterfactual densities based on linear quantile regression fits for several quantile levels and a simulation-based procedure. I.e., they generate counterfactual outcomes by using the estimated quantile regression coefficients and by drawing from the covariate distribution (to integrate-out the effect of the covariates) and by sampling quantile ranks from a uniform distribution. The resulting counterfactual densities need to be estimated by kernel density estimation. We consider our density regression-focused approach to be complementary to these existing approaches. This is particularly the case in settings with mixed-type dependent variables, as in our empirical application. Quantile-based methods are not applicable in such scenarios.

The issue of counterfactual density estimation has been studied in the existing literature.
\cite{dinardo1996labor} considers a similar decomposition of effects and a similar plug-in estimator for counterfactual densities. However, (i) they do not explicitly discuss causal implications of their estimated densities, (ii) they use kernel density estimates, which restricts the applicability in many settings and (iii) they consider differences between densities instead of ratios, which might be hard to interpret in low-density regions of the support. \cite{martinez2024counterfactual} model counterfactual densities using Kernel--Stein discrepancies, but rely on parametric assumptions. \cite{melnychuk2023normalizing} propose a deep learning method called `Interventional Normalizing Flows' for estimating fully parametric counterfactual densities. Similarly, \cite{kennedy2023semiparametric} approximate counterfactual densities by parametric models; for instance, relying on the exponential family, Gaussian mixture models, or on truncated series regression. Theoretically, they provide asymptotic results for the estimator, such as root-$n$ consistency and semiparametric efficiency bounds. They also consider `density effects', which they define based on distances or other measures of discrepancy. However, the above-mentioned approximations suffer from several limitations, namely potential model misspecification and the risk that the estimated densities can take negative values and do not integrate to one. Further, the specific setting of our empirical application with bimodalities and point masses cannot be adequately addressed by fixed parametric distributions. The focus of the present paper is therefore different from that of the above literature for two reasons. First, we explicitly embed our framework for causal inference on counterfactual densities within the decomposition methods literature. Second, while the above papers try to detect discrepancies between two densities in the form of scalar quantities, our method allows the localization of the regions in the support of the dependent variable where the discrepancies are substantial, by relying on ratios instead of distances.

Our contributions are fourfold. First, we propose a causal inference framework based on a multiplicative Oaxaca--Blinder decomposition of counterfactual densities. Compared with conventional additive decompositions, we thus focus on relative differences instead of absolute differences between densities, which has clear advantages for the analysis of low-density regions. A multiplicative approach further obeys the logic of Bayes Hilbert spaces, a suitable functional space for density functions. Second, we propose an estimation procedure that relies on a flexible additive model specification for the conditional densities that avoids both the restrictiveness of parametric models and the curse of dimensionality of fully nonparametric models. Further, the approach does not require continuous outcome variables but also allows for discrete and mixed-type distributions, as is the case in our empirical application. In practice, estimation can be carried out as an approximate Poisson regression problem. Third, the additive specification of the conditional densities allows us to further isolate the effect of different covariates on the covariate effect. Finally, on the empirical side, we use our counterfactual density methodology to gain more detailed insights into the East--West income gap in Germany that would otherwise have been lost when relying on mean-based procedures.

The remainder of this paper is structured as follows. Section~\ref{sec:counterfactual} introduces the model setup and notation and provides a definition of counterfactual densities as well as the corresponding counterfactual density effects. The Bayes Hilbert space approach for the estimation of conditional densities is discussed in Section~\ref{sec:estimation}. We provide a short simulation study to analyze the finite-sample properties of our estimation method in Section~\ref{sec:simulation}. In Section~\ref{sec:application} we employ our counterfactual density methodology to study the East--West income gap in Germany. Section~\ref{sec:conclusion} concludes.

\section{Counterfactual Densities and Causal Effects}\label{sec:counterfactual}

\subsection{Setup and Notation}\label{subsec:setup}

In this section, we introduce our framework for counterfactual densities. For this purpose, we will use the following notation. Let $Y\in\mathbb{R}$ be the dependent variable, $X\in\mathbb{R}^d$ the vector of covariates, and let $D\in\{0,1\}$ be the treatment variable, which takes the value one for the treated and zero otherwise. For the sake of exposition, we focus on binary treatment variables. However, extensions to more than two treatment groups will be discussed later. We consider the following potential outcome setting \citep{rubin1974estimating},
\begin{align*}
    Y=DY_1+(1-D)Y_0,
\end{align*}
where $Y_1$ is the potential outcome for the treated and $Y_0$ the potential outcome for the control group. Since only one potential outcome is observed, $Y_0$ is a latent variable for the treatment group and $Y_1$ for the control group.

Let $f_{Y_{\langle1,1\rangle}}$ and $f_{Y_{\langle0,0\rangle}}$ denote the unconditional densities of $Y_1$ and $Y_0$, respectively. Further define the conditional densities for treatment and control group as $f_{Y_1|X}(y|x)$ and $f_{Y_0|X}(y|x)$. In many applications, it may be of interest to define densities for the counterfactual quantities. For this purpose, $f_{Y_{\langle1,0\rangle}}$ denotes the (unconditional) counterfactual density of the treated if they had faced the covariate distribution of the control group,
\begin{align}\label{eq:counterfactualdensity}
    f_{Y_{\langle1,0\rangle}}(y):=\int_{\mathcal{X}_0} f_{Y_1|X}(y|x)dF_{X_0}(x),
\end{align}
where $F_{X_0}$ is the marginal cumulative distribution function (cdf) and $\mathcal{X}_0$ is the support of $X$ for the control group. Vice versa we can define $f_{Y_{\langle0,1\rangle}}$, the counterfactual density of the untreated if they had faced the covariate distribution of the treated group. These quantities are the object of study in \cite{dinardo1996labor}. \cite{chernozhukov2013inference} introduce a similar relationship for counterfactual distribution functions. Equation (\ref{eq:counterfactualdensity}) reveals that the counterfactual density is completely determined by the conditional density of the treated group and the marginal covariate distribution function of the control group. The crucial part in the estimation of counterfactual densities thus is to devise a suitable estimator of the conditional density. We will discuss this in detail in Section~\ref{sec:estimation}.

\subsection{Causal Inference and Counterfactual Density Effects}\label{sec:causal}

To attribute a causal interpretation to the counterfactual densities introduced in the previous subsection, we have to impose the following assumptions.

\begin{assumption}[Unconfoundedness]
Assume that $Y_0,Y_1\perp\!\!\!\perp D|X$.
\end{assumption}
\begin{assumption}[Overlap]
$0<P(D=1|X=x)<1$ for all $x\in\mathcal{X}$, where $\mathcal{X}$ denotes the support of $X$.
\end{assumption}
Both assumptions are commonly used in the causal inference literature (e.g., \cite{rosenbaum1983central}), and are not particular to our density-focused framework. However, they need to be carefully checked and discussed in any application of our framework.

A central issue is to define causal quantities which (i) are relevant for practitioners and (ii) are based on the counterfactual densities. One possibility is to consider the decomposition by \cite{oaxaca1973male} originally introduced for the mean, but studied by \cite{dinardo1996labor} in the context of densities and by \cite{chernozhukov2013inference} for distribution and quantile effects. The decomposition for densities is given by
\begin{align*}%\label{eq:decomposition_difference}
    f_{Y_{\langle1,1\rangle}}-f_{Y_{\langle0,0\rangle}}=\left[f_{Y_{\langle1,1\rangle}}-f_{Y_{\langle0,1\rangle}}\right]+\left[f_{Y_{\langle0,1\rangle}}-f_{Y_{\langle0,0\rangle}}\right].
\end{align*}
This gives three different effects, (i) the effect of changing the conditional density, (ii) the effect of changing the covariate distribution, and (iii) a combination of both. 

Recently, \cite{kennedy2023semiparametric} studied another quantity, which they labeled as `density effects'. Consider two densities, $g_1(y)$ and $g_0(y)$, and some discrepancy function $h:\mathbb{R}^2\to\mathbb{R}^{+}$. Then, their density effects are scalar quantities defined as
\begin{align*}
    \psi_h=D_h\{g_1(y),g_0(y)\}=\int h\left\{g_1(y),g_0(y)\right\}g_0(y)dy.
\end{align*}
Examples for $D_h$ include the total variation distance and the KL divergence. While \cite{kennedy2023semiparametric} did not explicitly consider a decomposition approach, their measure can easily be incorporated under an Oaxaca-Blinder type decomposition. Potential downsides of this approach are limited interpretability and relevance for practitioners, as well as the loss of information entailed by aggregating discrepancies between densities into a single number. Even though it might be a suitable measure to detect the existence of possible discrepancies, it does not provide any information on the direction of the effect. In general, focusing on a scalar quantity leads to a loss of a large proportion of the information. We argue that when studying counterfactual densities, the focus should lie on the analysis of the heterogeneity of possible effects, i.e., how the treatment affects certain regions of the distribution. E.g., the treatment can have an effect on the lower or upper tail of the distribution or it can affect the entire distribution with a shift in the mean. For example, the introduction of a minimum wage will primarily affect the lower tail of the income distribution.

Going into a similar direction as the decomposition of \cite{oaxaca1973male} and \cite{blinder1973wage}, we can consider an alternative decomposition of effects. Instead of looking at differences between densities, we consider ratios,
\begin{align}\label{eq:decomposition_ratio}
    \frac{f_{Y_{\langle1,1\rangle}}(y)}{f_{Y_{\langle0,0\rangle}}(y)}=\frac{f_{Y_{\langle1,1\rangle}}(y)}{f_{Y_{\langle0,1\rangle}}(y)}\times\frac{f_{Y_{\langle0,1\rangle}}(y)}{f_{Y_{\langle0,0\rangle}}(y)}.
\end{align}
The motivation for the use of a multiplicative decomposition is twofold.
First, for the estimation of conditional densities we rely on the use of Bayes Hilbert spaces, which are suitable spaces for density functions. These are vector spaces in which addition corresponds to multiplication and subtraction corresponds to taking ratios (for details see Section~\ref{sec:estimation}).
Related to this point, the use of ratios offers clear advantages over differences for the analysis of densities. Our proposal can detect discrepancies along the whole domain of the density, whereas differences can only find minor differences in  regions with low density values, e.g., tail regions can never show important differences in contrast to high density regions. Similar to \cite{chernozhukov2013inference}, we can define three kinds of effects.
\begin{itemize}[leftmargin=2cm]
    \item[Type 1.] The density effect of changing the conditional density (\emph{distribution effect}),
    \begin{align*}
        \operatorname{DE}(y):=\frac{f_{Y_{\langle1,1\rangle}}(y)}{f_{Y_{\langle0,1\rangle}}(y)}=\frac{\int_{\mathcal{X}_1} f_{Y_1|X}(y|x)dF_{X_1}(x)}{\int_{\mathcal{X}_1} f_{Y_0|X}(y|x)dF_{X_1}(x)}.
    \end{align*}
    \item[Type 2.] The density effect of changing the covariate distribution (\emph{covariate effect}),
    \begin{align*}
        \operatorname{CE}(y):=\frac{f_{Y_{\langle0,1\rangle}}(y)}{f_{Y_{\langle0,0\rangle}}(y)}=\frac{\int_{\mathcal{X}_1} f_{Y_0|X}(y|x)dF_{X_1}(x)}{\int_{\mathcal{X}_0} f_{Y_0|X}(y|x)dF_{X_0}(x)}.
    \end{align*}
    \item[Type 3.] The density effect of changing both the conditional density and the covariate distribution (\emph{total effect}),
    \begin{align*}
        \operatorname{TE}(y):=\frac{f_{Y_{\langle1,1\rangle}}(y)}{f_{Y_{\langle0,0\rangle}}(y)}=\frac{\int_{\mathcal{X}_1} f_{Y_1|X}(y|x)dF_{X_1}(x)}{\int_{\mathcal{X}_0} f_{Y_0|X}(y|x)dF_{X_0}(x)}.
    \end{align*}
\end{itemize}

\begin{remark}
For the sake of exposition, we currently focus on the special case of a binary treatment variable. But similar to \cite{chernozhukov2013inference}, the framework can easily be generalized to $K$ treatment groups and thus $K$ different potential outcomes. The corresponding counterfactual density, $f_{Y_{\langle k,l\rangle}}$, would for instance be based on the  conditional distribution of group $l$ and the covariate distribution of group $k$.
\end{remark}
\begin{remark}
While many real data applications have dependent variables with a continuous distribution, it is also worthwhile to study the case of discrete or mixed outcomes. For instance, income distributions  typically have a point mass at zero income.
\end{remark}

\subsection{Decomposing the Distribution and Covariate Effect}\label{subsec:decomposing}

An advantage of the classical mean-based Oaxaca-Blinder approach is the possibility to further decompose the distribution and covariate effects into individual contributions of covariates. Unfortunately, this advantage does not directly translate to the study of other nonlinear distributional quantities, such as quantile or distributional effects. \cite{rothe2015decomposing} demonstrates this problem for quantile treatment effects, even in the case of a linear model for the conditional quantiles. This is due to potential dependence among different covariates.
The same issue complicates any attempt to decompose our counterfactual density effects additively. As for quantile effects, the problem even persists in the case of a purely multiplicative density regression model as we will present in Section~\ref{sec:estimation}. A possible but not entirely satisfactory solution for this issue is the sequential conditioning approach of \cite{chernozhukov2013inference}, which suffers from the issue of path-dependence: in most applications the order of covariates is chosen arbitrarily. Further, \cite{rothe2012partial} argues that such an approach is unable to accurately reflect the impact of group differences in the marginal distribution of a single covariate.

Instead, we propose to isolate the contribution of individual covariates by considering marginal covariate effects, holding all remaining variables fixed at their control group distribution. This approach avoids path-dependence and has a transparent interpretation: each quantity measures the effect on the outcome density of shifting a single covariate's distribution from the control to the treatment group, while leaving the joint distribution of the remaining covariates unchanged. While the resulting quantities do not multiply to the total covariate effect, they nonetheless provide interpretable summaries of which covariates drive the overall covariate effect. 

Recall the definition of the covariate effect,
\begin{align*}
    \operatorname{CE}(y)%=\frac{f_{Y_{\langle0,1\rangle}}(y)}{f_{Y_{\langle0,0\rangle}}(y)}
    =\frac{\int f_{Y_0|X}(y|x)dF_{X_1}(x)}{\int f_{Y_0|X}(y|x)dF_{X_0}(x)}.
\end{align*}
Consider the marginal effect of the $j$-th variable on the density of $Y_k$, after integrating out the effect of all other variables using the marginal covariate distribution of group $l$,
\begin{align*}
    \widetilde{h}_{j,k|l}(y|x_j)=\int f_{Y_{k}|X}(y|x)dF_{X_{l,-j}}(x_{-j}),
\end{align*}
for $k,l=0,1$, where $F_{X_{l,-j}}$ denotes group $l$'s cdf of all covariates except variable $j$. Related quantities have been studied in the context of nonparametric estimation of the conditional mean, see e.g., \cite{linton1995kernel} and \cite{hardle2004nonparametric}. Then we can consider the following quantity to measure variable $X_j$'s contribution to the covariate effect, by changing its distribution (instead of that of all $X$) from control to treatment group,
\begin{align}\label{eq:CEj}
    \operatorname{CE}_j(y):=&\frac{\int f_{Y_0|X}(y|x)dF_{X_{0,-j}}(x_{-j})dF_{X_{1,j}}(x_j)}{\int f_{Y_0|X}(y|x)dF_{X_0}(x)} \nonumber \\ 
    =&\frac{\int \widetilde{h}_{j,0|0}(y|x_j)dF_{X_{1,j}}(x)}{\int \widetilde{h}_{j,0|0}(y|x_j)dF_{X_{0,j}}(x)}.
\end{align}
Under the additive model specification for the conditional density that we will introduce in Section~\ref{sec:estimation}, the partial effect $\widetilde{h}_{j,0|0}$ in (\ref{eq:CEj}) corresponds to the additive effect of variable $X_j$. The expression is therefore independent of the marginal distribution of $X_{-j}$ in this special case.
The quantity $\operatorname{CE}_j$ answers the following question: how would the outcome density change if only the marginal distribution of covariate $X_j$ were shifted from that of the control group to that of the treatment group, while all other covariates remain distributed as in the control group? It provides a direct answer to this question at each point $y$ of the support of the dependent variable, rather than compressing it into a scalar summary as in the mean case.

Complementary to \eqref{eq:CEj}, one can further analyze the distribution effect by looking at the contribution of a given covariate $X_j$ to the distribution effect. We define the contribution to the distribution effect as follows,
\begin{align}\label{eq:DEj}
    \operatorname{DE}_j(y)&:=\frac{\int f_{Y_1|X}(y|x)dF_{X_{1,-j}(x_{-j})}dF_{X_{0,j}(x_j)}}{\int f_{Y_0|X}(y|x)dF_{X_0}(x)}\nonumber\\
    &=\frac{\int\widetilde{h}_{j,1|1}(y|x_j)dF_{X_{0,j}}(x_j)}{\int \widetilde{h}_{j,0|0}(y|x_j)dF_{X_{0,j}}(x)}.
\end{align}

The quantity $\operatorname{DE}_j$ compares the partial effect of covariate $X_j$ on the conditional density across the two groups. Specifically, the numerator evaluates the treatment group's structural relationship between $X_j$ and $Y$, while the denominator evaluates the control group's, both integrated over the common (control group) distribution of $X_j$. Thus, $\operatorname{DE}_j(y)$ measures how much the density at $y$ would change if only the way $X_j$ shapes the conditional density were switched from the control to the treatment group specification, while the distribution of $X_j$ itself is held fixed. This isolates the role of $X_j$ in driving the distribution effect, and can be a useful tool for identifying the set of variables that are most responsible for the distribution effect.

\section{Counterfactual Density Estimation}\label{sec:estimation}

In this section, we present our estimation procedure for the counterfactual densities. The critical step in the estimation is to obtain a suitable estimate of the conditional densities. Let $\{(Y_{ki},X_{ki})\}_{i=1}^{n_k}$ denote a sample of $n_k$ i.i.d. copies of $(Y_k,X_k)$, for $k=0,1$. We consider a plug-in estimator of the form
\begin{align*}
    \widehat{f}_{Y_{\langle1,0\rangle}}(y):=\int_{\mathcal{X}_0} \widehat{f}_{Y_1|X}(y|x)d\widehat{F}_{X_0}(x),
\end{align*}
where $\widehat{f}_{Y_1|X}(y|x)$ is an estimator of the conditional density and
\begin{align*}
    \widehat{F}_{X_0}(x)=\frac{1}{n_0}\sum_{i=1}^{n_0}\mathbf{1}_{\{X_{0i}\leq x\}}
\end{align*}
is the empirical distribution function of $X_0$, with the inequality $X_{0i}\leq x$ interpreted entry-wise for each $x_j$. $\widehat{f}_{Y_{\langle0,1\rangle}}(y)$ and $\widehat{F}_{X_1}(x)$ can be defined analogously. \cite{dinardo1996labor} propose a kernel density estimation procedure for the conditional densities. However, such fully nonparametric methods suffer from the curse of dimensionality and do not work well in settings with moderate to large dimensions of covariates. Other approaches suffer from restrictive parametric assumptions or do not provide suitable estimates for densities, i.e., the non-negativity or integrating-to-one constraints might be violated. Instead, for the estimation of the conditional densities we follow the Bayes Hilbert space approach of \citet{maier2024conditional}, who developed a flexible additive model framework for modeling conditional densities.

\subsection{Bayes Hilbert Spaces}\label{subsec:bayes_space}

Before setting up the density regression model, we first provide a concise introduction to the used Bayes Hilbert spaces \citep{boogart2010bayes,boogart2014bayes}. For a more detailed introduction we refer the reader to \citet{maier2025additive} and \citet{maier2024conditional}. The \emph{Bayes Hilbert space} on the measurable space $(\mathcal{T},\mathcal{A})$ with reference measure $\mu$ is defined by $B^2(\mu)=B^2(\mathcal{T},\mathcal{A},\mu):=\{f\in B(\mu)|\int_{\mathcal{T}}(\log f)^2d\mu<\infty\}$, where $B(\mu)$ is a \emph{Bayes space} (i.e., a set of equivalence classes of $\mu$-densities that are $\mu$-a.e. positive and unique) with reference measure $\mu$. This is a vector space with addition, $f_1\oplus f_2=_{\mathcal{B}}f_1f_2$, and scalar multiplication, $\alpha\odot f_1=_{\mathcal{B}}(f_1)^{\alpha}$, for $f_1,f_2\in B^2(\mu)$ and $\alpha\in\mathbb{R}$, where $=_{\mathcal{B}}$ denotes equality up to scale. A crucial concept in this framework is the \emph{centered log-ratio (clr)} transformation, $\operatorname{clr}(f):=\log(f)-\frac{1}{\mu(\mathcal{T})}\int_{\mathcal{T}}\log(f)d\mu$, which maps an element in $B^2(\mu)$ to a function in $L^2_0(\mu)=L_0^{2}(\mathcal{T},\mathcal{A},\mu):=\{\tilde{f}\in L_0^2(\mu)|\int_{\mathcal{T}}\tilde{f}d\mu=0\}$, a closed subspace of $L^2(\mu)$. Transforming the data is beneficial for practical implementation and computational reasons as it enables the use of tools established for $L^2$-spaces. The clr transform is an isometric isomorphism, and it is bijective with inverse transformation $\operatorname{clr}^{-1}(\tilde{f})=_{\mathcal{B}}\exp(\tilde{f})$. $B^2(\mu)$ is a Hilbert space, with inner product $\langle f_1,f_2\rangle_{B^2(\mu)}=\int_{\mathcal{T}}\operatorname{clr}(f_1)\cdot\operatorname{clr}(f_2)d\mu$. Although Bayes Hilbert spaces are defined for arbitrary measurable spaces, we will focus on $\mathcal{T}\subset\mathbb{R}$. Following \citet{maier2024conditional}, we distinguish between three cases. For the \emph{continuous} case, we have $\mathcal{T}=[a,b]$ and $\mu$ is the Lebesgue measure $\lambda$ on $\mathcal{T}$. For the \emph{discrete} case, $\mathcal{T}=\{t_1,\ldots,t_D\}$ and $\mu$ is a weighted sum $\delta$ of Dirac measures. Finally, we consider the \emph{mixed} case with $\mathcal{T}=[a,b]\cup\{t_1,\ldots,t_D\}$ and $\mu=\lambda+\delta$. To demonstrate the importance of the latter case, note that in our empirical application we will look at income distributions in Germany, which have mixed-type densities with additional point masses at zero and for incomes above a certain threshold. 

\subsection{Additive Density Regression in Bayes Hilbert Spaces}\label{subsec:bayes_regression}

We now present the flexible additive density regression setup of \citet{maier2024conditional}. In the following, we suppress the group-specific index of the data for the sake of exposition. Consider an i.i.d. sample of observations, $(y_i,x_i)\in\mathcal{T}\times\mathcal{X}$, $\mathcal{X}\subseteq\mathbb{R}^d$, $i=1,\ldots,n$. The conditional density of $Y$ given $X=x_i$, denoted by $f_i:=f(Y|X=x_i)$, is assumed to be an element of the Bayes Hilbert space $B^2(\mu)$. As described in the previous subsection, the framework is flexible enough to handle continuous, discrete, as well as mixed distributions and data. We assume the following additive structure,
\begin{align}\label{eq:density}
    f_i=\bigoplus_{j=1}^{J}h_j(x_i),
\end{align}
where the partial effects $h_j$ are also elements of the Bayes Hilbert space, $h_j(x_i)\in B^2(\mu)$. Each partial effect can depend on one, several (for interactions) or no covariates (the intercept) and can be linear or nonlinear in $x_i$. Each effect is assumed to be represented by the following tensor product basis,
\begin{align}\label{eq:basis}
    h_j(x)=\bigoplus_{l=1}^{d_j}\bigoplus_{m=1}^{d_{\tau}}\theta_{j,l,m}\odot b_{j,l}(x)\odot b_{\mathcal{T},m},
\end{align}
where $b_{j,l}:\mathbb{R}^{d}\to\mathbb{R}$, $b_{\mathcal{T},m}\in B^2(\mu)$ are basis functions over the covariates and over $\mathcal{T}$, respectively, and $\theta_{j,l,m}\in\mathbb{R}$ are the corresponding coefficients. For identification, we center smooth main effects around the intercept $\beta_0$ and interactions around corresponding main effects, see \citet{maier2024conditional}.

\begin{remark}\label{remark:decomposition}
Under model specification (\ref{eq:density}), and if the partial effects contain no interaction terms, we can rewrite the decomposition of the covariate effect introduced in (\ref{eq:CEj}),
\begin{align*}
    \operatorname{CE}_j(y)&=\frac{\int f_{Y_0|X}(y|x)dF_{X_{0,-j}}(x_{-j})dF_{1,j}(x_j)}{\int f_{Y_0|X}(y|x)dF_{X_0}(x)}\\
    &=\frac{\int h_{0,j(y,x_j)}dF_{1,j}(x_j)}{\int h_{0,j(y,x_j)}dF_{0,j}(x_j)}.
\end{align*}
\end{remark}

Applying the centered log-ratio (clr) transformation to (\ref{eq:density}) and using the basis function representation in (\ref{eq:basis}) yields
\begin{align*}
    \tilde{f}_i&=\operatorname{clr}\left(f_i\right)=\sum_{j=1}^J\sum_{l=1}^{d_j}\sum_{m=1}^{d_{\mathcal{T}}}\theta_{j,l,m}b_{j,l}(x_i)\tilde{b}_{\mathcal{T},m}\\
&=\left(b(x_i)\otimes \tilde{b}_{\mathcal{T}}\right)^{\top}\boldsymbol{\theta},
\end{align*}
where $\tilde{b}_{\mathcal{T},m}=\operatorname{clr}(b_{\mathcal{T},m})$, $b(x)=(b_{1,1}(x),\ldots,b_{J,d_J}(x))^{\top}\in\mathbb{R}^{\sum_{j=1}^{J}d_j}$, and $\tilde{b}_{\mathcal{T}}=(\tilde{b}_1,\ldots,\tilde{b}_{d_{\mathcal{T}}})^{\top}\in(L_0^2(\mu))^{d_{\mathcal{T}}}$ are vectors of basis functions, and the corresponding parameter vector $\boldsymbol{\theta}=(\boldsymbol{\theta}_1^{\top},\ldots,\boldsymbol{\theta}_J^{\top})^{\top}\in\mathbb{R}^R$ with $\boldsymbol{\theta}_j=(\theta_{j,1,1},\ldots,\theta_{j,d_j,d_{\mathcal{T}}})^{\top}$ and the dimension of $\boldsymbol{\theta}$ is $R=\sum_{j=1}^J d_jd_{\mathcal{T}}$. The choice of the covariate-specific basis functions $b_j(x_i)$ depends on the type of the considered effect. For instance, a smooth non-linear effect can be modeled via B-splines. Similarly, the choice of the basis functions $\tilde{b}_{\mathcal{T}}$ depends on the reference measure $\mu$. \citet{maier2025additive} describe constructions based on transformations of B-splines and of indicators for the continuous and the discrete part, respectively.

In principle, $\boldsymbol{\theta}$ can be estimated via maximum likelihood estimation, with likelihood and log-likelihood functions given by
\begin{align*}
    L(\boldsymbol{\theta})&=\prod_{i=1}^n\frac{\exp\left(\tilde{f}_i\right)}{\int_{\mathcal{T}}\exp\left(\tilde{f}_i\right)d\mu},\\
    \ell(\boldsymbol{\theta})&=\sum_{i=1}^{n}\left(\tilde{f}_i-\log\int_{\mathcal{T}}\exp(\tilde{f}_i)d\mu\right)\\
&=\sum_{i=1}^{n}\left(b(x_i)\otimes\tilde{b}_{\mathcal{T}}(y_i)\right)^{\top}\boldsymbol{\theta}-\log\int_{\mathcal{T}}\exp\left[\left(b(x_i)\otimes\tilde{b}_{\mathcal{T}}\right)^{\top}\boldsymbol{\theta}\right]d\mu.
\end{align*}

To increase the smoothness of the estimated densities, it is also possible to include additional penalty terms for the coefficients, and thus to consider a penalized log-likelihood. The estimation can be computationally challenging due to the presence of the integral term in the log-likelihood. Following \citet{maier2024conditional}, we instead estimate $\boldsymbol{\theta}$ by approximating the problem via additive Poisson regression.

\subsection{Estimation via Multinomial and Poisson Regression}\label{subsec:estimation}

For reducing the computational burden, the estimation procedure can be approximated by using a (shifted) multinomial log-likelihood. In this section we focus on the continuous case for notational simplicity, although \citet{maier2024conditional} show that the discrete and mixed cases can also be covered. For this purpose, we need to partition the support of $Y$ into discrete histogram bins. Let $a=a_0<a_1<\ldots<a_G=b$. Then $U_g=[a_{g-1},a_g)$ for $g=1,\ldots,G-1$ and $U_G=[a_{G-1},a_G]$ partition the interval $[a,b]$. The values of the histogram are $n_g^i=\mathbf{1}_{\{y_i\in U_g\}}$ and the corresponding histogram widths are $\Delta_g=a_g-a_{g-1}$ for $g=1,\ldots,G$. Further, denote the bin center of histogram bin $U_g$ as $u_g$. The vector $(n_1^i,\ldots,n_G^i)$ can be viewed as a realization of a multinomial variable with sample size $1$ and with class probabilities
\begin{align*}
    p_g^i(\boldsymbol{\theta})=\frac{\Delta_g\exp\left[\left(b(x_i)\otimes\tilde{b}_{\mathcal{T}}(u_g)\right)^{\top}\boldsymbol{\theta}\right]}{\sum_{k=1}^G\Delta_k\exp\left[\left(b(x_i)\otimes\tilde{b}_{\mathcal{T}}(u_k)\right)^{\top}\boldsymbol{\theta}\right]}.
\end{align*}
The multinomial log-likelihood up to constants is
\begin{align*}
    \ell^{mn}(\boldsymbol{\theta})\propto \sum_{i=1}^n\sum_{g=1}^G\left(\left(b(x_i)\otimes\tilde{b}_{\mathcal{T}}(u_g)\right)^{\top}\boldsymbol{\theta}-\log\sum_{k=1}^G\Delta_k\exp\left[\left(b(x_i)\otimes\tilde{b}_{\mathcal{T}}(u_k)\right)^{\top}\boldsymbol{\theta}\right]\right).
\end{align*}
\citet{maier2024conditional} show that the multinomial log-likelihood converges to the Bayes Hilbert space log-likelihood as the maximal bin size approaches zero, as well as the convergence of the corresponding maximum likelihood estimator and the inverse Fisher information used for inference. Further, they show the equivalence of the multinomial and a certain Poisson likelihood. In particular, for computational reasons it is beneficial to pool observations that share the same combination of covariates, and fit the histogram counts using a Poisson model with an additional intercept parameter for each unique covariate combination. We can thus rely on additive Poisson regression for estimation of the parameter vector $\boldsymbol{\theta}$.

\subsection{Uncertainty Quantification}\label{subsec:uncertainty}

Having introduced the additive density regression framework and the associated estimation procedure, a natural next question concerns the issue of uncertainty quantification. In particular, it is of major interest to empirical researchers whether the distribution and covariate effects are significant, i.e., different from one. For this purpose, we rely on the asymptotic results of \cite{maier2024conditional}. We propose drawing values $\boldsymbol{\theta}_k^{(b)}$, $b=1,\ldots,B$, of the regression parameters from the $(1-\alpha)$ Wald confidence regions defined in Lemma A.15 of the above paper, for a given significance level $\alpha$. For each simulation iteration, we obtain the corresponding conditional density based on the simulated parameter values and the fixed basis functions, $\widehat{f}_{Y_k|X}^{(b)}(y|x)=(b(x)\otimes \tilde{b}_{\mathcal{T}})^{\top}\boldsymbol{\theta^{(b)}_k}$, for $k=0,1$. Further, we calculate the respective counterfactual densities by integrating with respect to the covariate distributions of the treatment and control groups, $\widehat{f}^{(b)}_{Y_{\langle k,l\rangle}}(y)=\int_{\mathcal{X}_l} \widehat{f}^{(b)}_{Y_k|X}(y|x)d\widehat{F}_{X_l}(x)$, for $k,l=0,1$. Since the conditional densities of both groups are estimated independently, we can follow the above procedure separately for each group. We thus obtain estimates of the respective density effects based on simulations from the asymptotic confidence regions of the model parameters. These estimates can be plotted alongside the original estimate of the respective density effect and thus serve to quantify estimation uncertainty.

\section{Simulation Study}\label{sec:simulation}

\subsection{Simulation Setup}

In this section, we study the finite sample performance of our proposed estimator for the counterfactual densities using a simulation study. To enable the comparison with an alternative estimator for the conditional densities, we restrict our attention on a data-generating process with categorical covariates and continuous dependent variables. We consider a setting with $d=3$ covariates, each taking two possible values. We assume the additive model specification for the conditional densities introduced in \eqref{eq:density}. The partial effects $h_j$ are generated as beta density functions with different parameter values, $h_j(x)=x_j\odot \beta_j$, $j=1,2,3$. As a consequence, the conditional densities are also beta densities. We consider sample sizes ranging from $n=500$ to $n=100{,}000$. Further details on the data-generating process and other aspects of the simulation study are provided in Section~\ref{sec:appendix_sim} of the supplementary material. It should be noted that our model is not correctly specified, as the spline basis functions only approximate the true conditional densities.

The performance is evaluated by the total variation (TV) distance between the true and estimated counterfactual densities. We further evaluate the estimation accuracy of the conditional densities using the same metric. To provide a benchmark, we compare the estimation accuracy with an alternative approach based on kernel density estimation with a Gaussian kernel and Silverman's rule of thumb bandwidth, carried out separately for every covariate combination, which is possible in this case with only binary covariates. All simulation results are based on $1{,}000$ Monte Carlo iterations.

\subsection{Simulation Results}

Table \ref{table:simulation_counterfactual} reports the estimation accuracy of the four counterfactual densities in terms of the TV distance between estimated and true densities. As expected, the estimation becomes more accurate with increasing sample size. The table further reports the estimation accuracy of the benchmark method based on kernel density estimates of the conditional densities. The performance of both methods is quite similar in small samples, whereas the Bayes Hilbert space approach has a slight advantage in settings with larger samples ($n=10{,000},20{,}000$). However, for the largest sample size ($n=100{,}000$), the performance of the two methods appears to converge.

Since this is the key step of our estimation methodology, we separately analyze the estimation accuracy of the conditional densities using the TV distance between the true and estimated densities, comparing the Bayes–Hilbert approach with kernel density estimation. To make the estimation results easier to interpret, we aggregate the results by averaging the TV distances over all covariate combinations. Interestingly, the results in Table~\ref{table:simulation_conditional} tell a different story from the estimation results of the unconditional counterfactual densities. In fact, the estimation accuracy of the Bayes Hilbert space approach is higher in all the settings we consider. The reason for this is that the method makes explicit use of the multiplicative structure of the conditional densities. To illustrate this point, for the estimation of one particular conditional density, $f_{Y_j|X}(y|X=x)$, the Bayes Hilbert space approach can borrow strength from data that do not belong to this conditional density. I.e., an observation $i$ with $X_i\neq x$ can still impact the estimation of the conditional density. In contrast, the kernel density estimator can only make use of observations for which $X_i=x$. However, since the estimation of the counterfactual densities involves taking averages over the estimated conditional densities, the kernel density estimator has the advantage that the estimates of the different conditional densities are statistically independent. Therefore, the results for the counterfactual densities in Table~\ref{table:simulation_counterfactual} are less clear than the results in Table~\ref{table:simulation_conditional}.

We want to point out that for comparison purposes the simulation study is restricted to discrete regressors and a continuous outcome variable. An additional advantage of the Bayes Hilbert space approach is that it can be easily applied to settings with continuous regressors as well as discrete and mixed-type outcome variables. To account for continuous regressors, the kernel density benchmark would also need to rely on smoothing in the covariate dimension, which would make the estimation problem of the conditional densities much more difficult. In contrast, the Bayes Hilbert space approach can be readily applied in these settings as well.

\begin{table}[tbp]
\centering
\begin{tabular}{l|rrrr|rrrr}
	\hline 
   \multicolumn{1}{l}{} & \multicolumn{4}{c}{Bayes Hilbert} & \multicolumn{4}{c}{Kernel density} \\
  $n$ & $f_{Y_{\langle1,1\rangle}}$ & $f_{Y_{\langle1,0\rangle}}$ & $f_{Y_{\langle0,1\rangle}}$ & $f_{Y_{\langle0,0\rangle}}$ & $f_{Y_{\langle1,1\rangle}}$ & $f_{Y_{\langle1,0\rangle}}$ & $f_{Y_{\langle0,1\rangle}}$ & $f_{Y_{\langle0,0\rangle}}$ \\ 
  \hline
  500 & 0.036 & 0.042 & 0.049 & 0.048 & 0.036 & 0.040 & 0.052 & 0.048 \\
  1{,}000 & 0.028 & 0.032 & 0.038 & 0.037 & 0.028 & 0.032 & 0.039 & 0.036 \\ 
  5{,}000 & 0.015 & 0.017 & 0.021 & 0.019 & 0.016 & 0.019 & 0.021 & 0.021 \\ 
  10{,}000 & 0.011 & 0.013 & 0.015 & 0.014 & 0.013 & 0.016 & 0.016 & 0.017 \\
  20{,}000 & 0.009 & 0.010 & 0.012 & 0.011 & 0.010 & 0.013 & 0.012 & 0.013 \\ 
  100{,}000 & 0.006 & 0.006 & 0.008 & 0.008 & 0.006 & 0.008 & 0.007 & 0.008 \\ 
 \hline 
\end{tabular}
\caption{Total variation distance between estimated and true counterfactual densities. The results on the left side of the table are based on estimated conditional densities using the Bayes Hilbert space approach; the results on the right side are based on kernel density estimation.}
\label{table:simulation_counterfactual}
\end{table}

\begin{table}[tbp]
\centering
\begin{tabular}{l|rr|rr}
	\hline 
  \multicolumn{1}{l}{} & \multicolumn{2}{c}{Bayes Hilbert} & \multicolumn{2}{c}{Kernel density} \\
  $n$ & $f_{Y_1|X}$ & $f_{Y_0|X}$ & $f_{Y_1|X}$ & $f_{Y_0|X}$ \\ 
  \hline
  500 & 0.064 & 0.073 & 0.086 & 0.096 \\
  1{,}000 & 0.047 & 0.056 & 0.066 & 0.074 \\ 
  5{,}000 & 0.025 & 0.032 & 0.037 & 0.041 \\ 
  10{,}000 & 0.019 & 0.024 & 0.029 & 0.032 \\ 
  20{,}000 & 0.015 & 0.018 & 0.022 & 0.025 \\ 
  100{,}000 & 0.008 & 0.011 & 0.013 & 0.014 \\ 
 \hline 
\end{tabular}
\caption{Average total variation distance between estimated and true conditional densities (averaged over all conditional densities). The results on the left side of the table are based on estimated conditional densities using the Bayes Hilbert space approach; the results on the right side are based on kernel density estimation.}
\label{table:simulation_conditional}
\end{table}

\section{Decomposing the East--West Income Gap in Germany}\label{sec:application}

\subsection{Data and Background}\label{subsec:application_data}

We apply our counterfactual density framework to analyze the East--West gap in gross incomes in Germany. Most existing studies on this subject consider a location-based definition of Easterners and Westerners, irrespective of the place of birth and socialization \citep{burda1997getting,kluge2018decomposing}. However, this approach is likely to suffer from endogeneity bias. Place of residence and place of work are to a large extent choice variables that might depend on latent factors, thereby distorting the effects of interest. In contrast to these above studies, \citet{dickey2021persistent} are interested both in a location-based and an origin-based East--West gap.
These existing papers use Oaxaca--Blinder type decomposition methods, which are based either on mean regression, quantile regression, or the unconditional quantile regression approach of \citet{fortin2011decomposition}. We use our decomposition approach for densities, which allows to look at effects on the whole distribution in an interpretable way while also being able to cover zero as well as positive incomes.

Due to the aforementioned endogeneity issues, we restrict our analysis to an origin-based definition of the East--West gap. The data are obtained from the Socio-Economic Panel (SOEP), which provides person-specific information on demographic and socio-economic aspects (see \citet{goebel2019german}). We identify Easterners using the variable `loc1989', which provides information on the place of residence immediately before the fall of the Berlin Wall and German reunification. If this information is not available, we further classify a person as an Easterner if their birth region lies geographically in the East. This secondary identification is essential for categorizing individuals born after 1989. Formally, this is done using the variable `birthregion\_ew'. We restrict our analysis to individuals aged 18–67. We further exclude retirees, students (both at school and at university), as well as persons with disabilities. To account for over- and under-representation of certain demographic groups, we use the corresponding cross-sectional individual weights provided in the dataset.

The dependent variable is monthly gross labor income, which includes both primary and secondary income sources. To ensure comparability across time, all income values are adjusted to 2021 price levels. We note that the density of this variable is of a mixed type, with a point mass at zero. Additionally, we use an upper bound for the support of the income variable at EUR $10{,}000$, setting values above to the category $10{,}000+$ represented by a second point mass. This introduces another point mass at the threshold value. As covariates, we consider sex ($x_{sex}$); a categorical variable for level of education ($x_{edu}$); an indicator variable for whether the place of residence is urban or rural ($x_{rural}$); and a categorical variable for employer size ($x_{size}$). Apart from these categorical explanatory variables, we also consider the age of the individual as a smooth effect. By the logic of the Oaxaca--Blinder approach, we run two separate regressions for Easterners and Westerners, with the following additive density regression model
\begin{align*}
    f_{Y_k|X}(y|x)&=\beta_{0,k}(y)\oplus\beta_{edu,k}(y,x_{edu})\oplus\beta_{rural,k}(y,x_{rural})\\&\oplus\beta_{size,k}(y,x_{size})\oplus\beta_{sex,k}(y,x_{sex})\oplus g_k(y,x_{age}),
\end{align*}
where $k=1$ for the East and $k=0$ for the West. The partial effects $\beta_{j,k}$ denote group-specific intercepts for the $j$-th categorical variables, where $\beta_{j,k}=0_{B^2}$, the additive neutral element of the Bayes Hilbert space for the respective reference category, and $g_k(\cdot)$ represents a smooth effect for age. Since East Germans are our treatment group and West Germans the control group, the distribution and covariate effects are defined as $\operatorname{DE}(y)=f_{Y_{\langle\text{East},\text{East}\rangle}}(y)/f_{Y_{\langle\text{West},\text{East}\rangle}}(y)$ and $\operatorname{CE}(y)=f_{Y_{\langle\text{West},\text{East}\rangle}}(y)/f_{Y_{\langle\text{West},\text{West}\rangle}}(y)$, respectively. To account for changes over time, we conduct separate analyses for the years 1991 ($n_0=3{,}132$, $n_1=6{,}764$), 2001 ($n_0=3{,}936$, $n_1=9{,}929$), 2011 ($n_0=4{,}893$, $n_1=12{,781}$) and 2021 ($n_0=2{,}447$, $n_1=6{,}433$). For the years 1991 and 2001, we choose smaller upper bounds for the continuous part of the income distribution: EUR $5{,}000$ and EUR $7{,}000$, respectively.

\subsection{Estimation Results}\label{subsec:application_estimation}

\begin{figure}[tbp]
\centering
\includegraphics[width=0.9\textwidth]{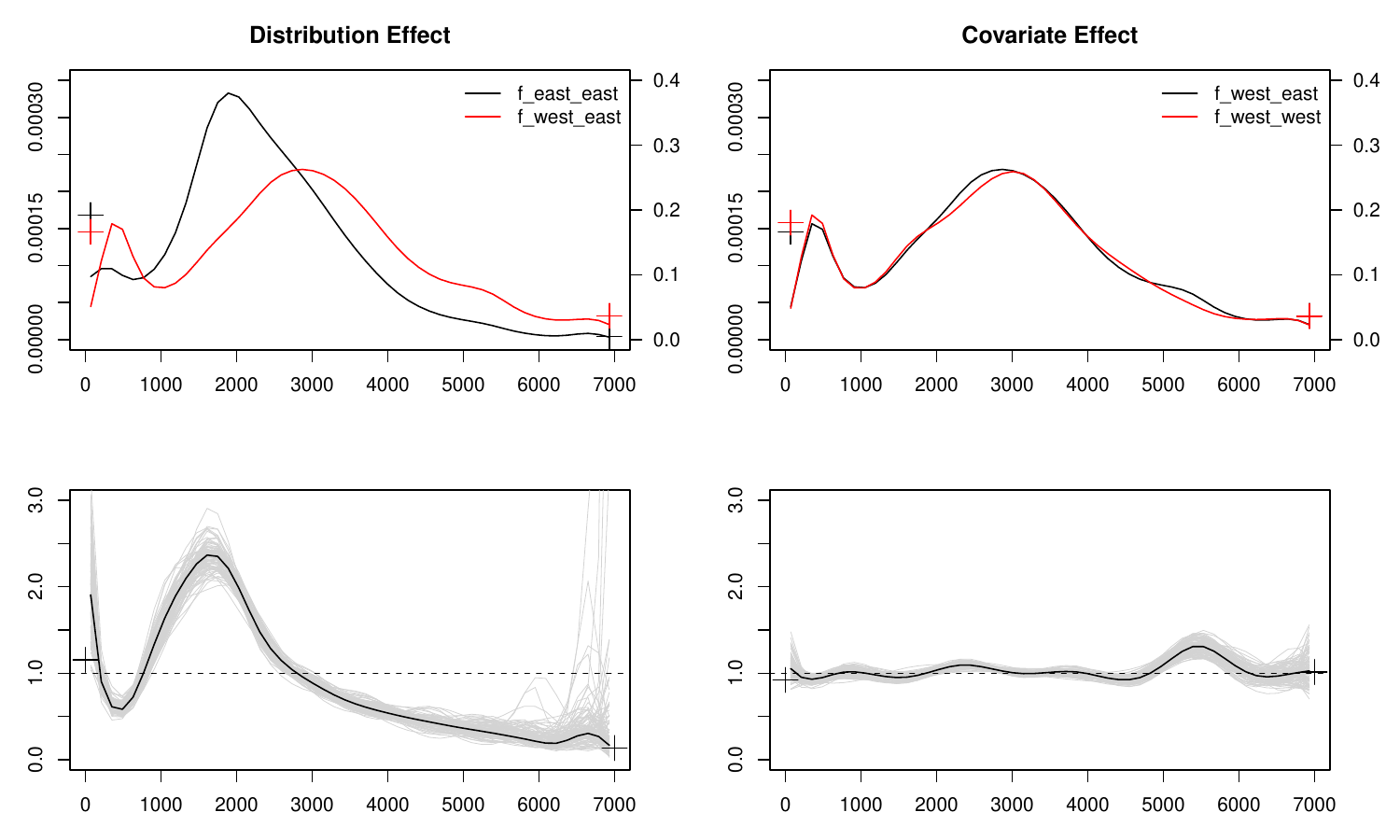}
\caption{Decomposition of the total density effect into distribution effect (left panels) and covariate effect (right panels) for the year 2001. The top panels show the estimated counterfactual densities, $f_{Y_{\langle\text{East},\text{East}\rangle}}$, $f_{Y_{\langle\text{West},\text{East}\rangle}}$, and $f_{Y_{\langle\text{West},\text{West}\rangle}}$. The lower panels show the estimated density effects $\operatorname{DE}(y)$ and $\operatorname{CE}(y)$, with 100 draws from the 95\% confidence region.}
\label{fig:counter01}
\end{figure}

\begin{figure}[tbp]
\centering
\includegraphics[width=0.9\textwidth]{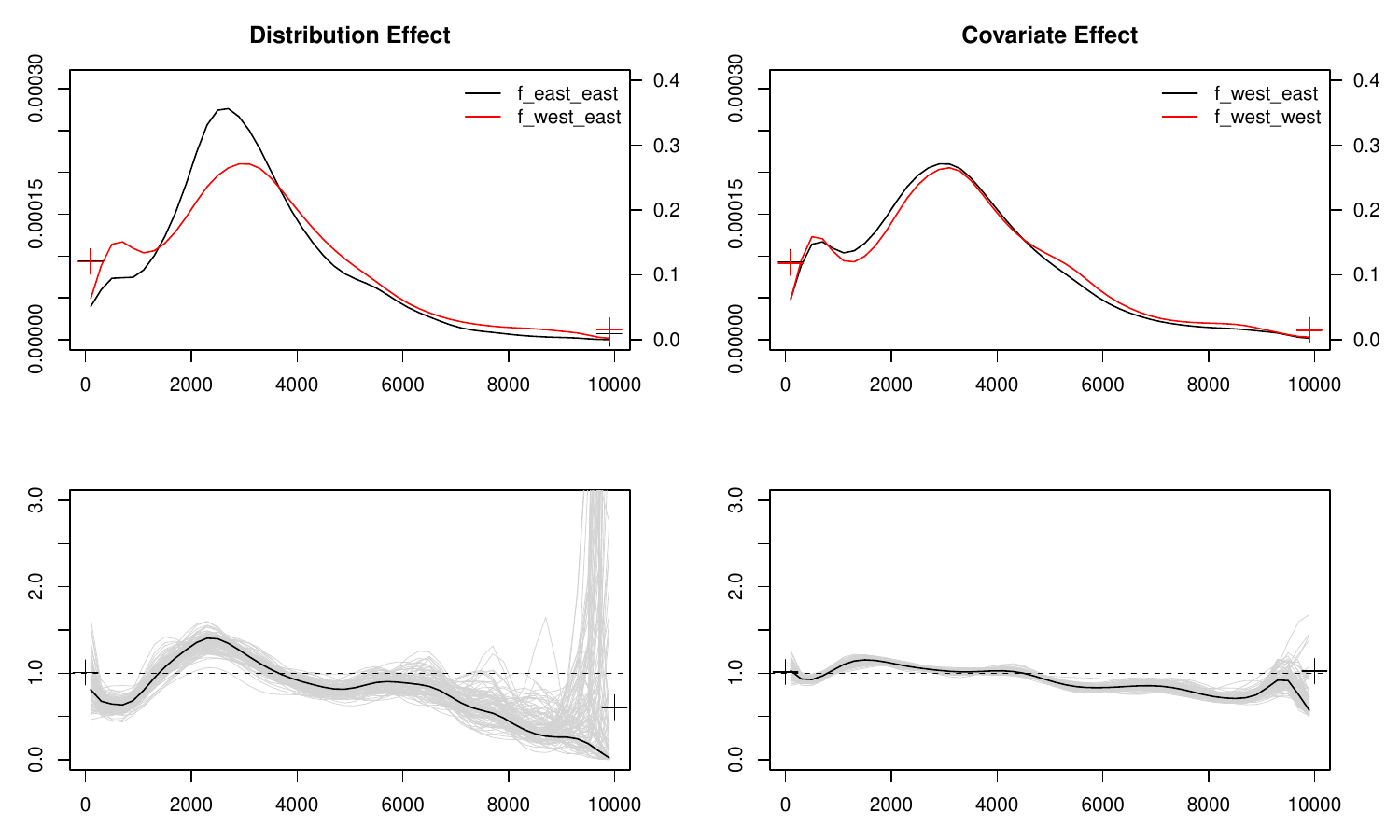}
\caption{Decomposition of the total density effect into distribution effect (left panels) and covariate effect (right panels) for the year 2021. The top panels show the estimated counterfactual densities, $f_{Y_{\langle\text{East},\text{East}\rangle}}$, $f_{Y_{\langle\text{West},\text{East}\rangle}}$, and $f_{Y_{\langle\text{West},\text{West}\rangle}}$. The lower panels show the estimated density effects $\operatorname{DE}(y)$ and $\operatorname{CE}(y)$, with 100 draws from the 95\% confidence region.}
\label{fig:counter21}
\end{figure}

This section presents the estimation results for the counterfactual densities, the corresponding distribution and covariate effects, and their development over time.
The results for the East--West gap for 2001 and 2021 are visualized in Figures \ref{fig:counter01} and \ref{fig:counter21}. We refer to Figures \ref{fig:counter91} and \ref{fig:counter11} in Section~\ref{sec:application_other} of the supplementary material for additional results for 1991 and 2011. We add 100 draws from the 95\% confidence regions for the distribution and covariate effects to quantify the estimation uncertainty, following the procedure outlined in Subsection~\ref{subsec:uncertainty}. First, we observe that the discrepancies between income densities for East and West have decreased substantially over recent decades. This confirms both previous empirical findings and theoretical predictions about East--West wage convergence. Still, the distribution effect for 2021 shows that differences persist in the lower as well as upper regions of the income distribution. For instance, even if West Germans were to have the same age structure and demographic characteristics as East Germans, they would be far more likely to be in the upper tail of the income distribution. As a second major empirical result, we find that the density-based East--West gap can be overwhelmingly attributed to the distribution effect, with the covariate effect only playing a minor role. Thus, the observed differences can not be well explained by factors such as education or urbanization.

To illustrate the benefit of our density-focused analysis, we include a comparison with a classical (additive) Oaxaca--Blinder decomposition for the mean incomes of the treatment (East) and control (West) groups. When looking at the left side of Table~\ref{table:oaxaca_mean_distance}, it becomes clear that a mean-based analysis provides a simple, scalar summary that is straightforward to interpret. However, this interpretability comes at the expense of not being able to capture the nuances of the differences. For example, the higher average income of Westerners can stem either from a large share of high incomes or from a very low share of low incomes. Empirically, the results reaffirm our finding that the covariate effect is dominated by the distribution effect, and that the latter effect is decreasing over time. Interestingly, the covariate effect is positive in the years following the reunification, but the sign switches between the years 2001 and 2011.

We also compare our results with the `density effects' proposed by \cite{kennedy2023semiparametric}, which are again a scalar measure for the discrepancy between two (counterfactual) densities. As a metric we choose the total variation distance, and we also decompose the effect additively into a distribution and covariate effect. For simplicity, we use the same estimated counterfactual densities as in our main analysis. The results on the right hand side of Table~\ref{table:oaxaca_mean_distance} show the fundamental limitation of this approach. Even if the approach can reliably detect discrepancies between the estimated densities beyond the case of simple mean shifts, it is still incapable of identifying the regions of the income distributions that are responsible for the discrepancy. An additional disadvantage is the lack of information about the direction of the effect across the distribution. This information is of course of utmost importance for policymakers. We therefore argue that scalar measures for discrepancies between densities can be beneficial as an additional tool for analysis, but important information about the direction and location of the effect is lost in the process. 

\begin{table}[tbp]
\centering
\begin{tabular}{l|rr|rr}
	\hline 
  %\multicolumn{1}{l}{} & \multicolumn{2}{r}{Mean decomposition} & \multicolumn{2}{r}{TV-distance decomposition} \\
  & DE & CE & DE & CE \\ 
  \hline
  1991 & --1426 & 349 & 0.558 & 0.072 \\ 
  2001 & --703 & 102 & 0.225 & 0.025 \\ 
  2011 & --481 & --71 & 0.149 & 0.041 \\ 
  2021 & --213 & --145 & 0.097 & 0.034 \\  
 \hline 
\end{tabular}
\caption{Results for the Oaxaca--Blinder type decomposition, i.e., the distribution effect (DE) and the covariate effect (CE), for mean differences (left side of the table) and analysis of the total variation distance between counterfactual densities (right side of the table) following \citet{kennedy2023semiparametric}.}
\label{table:oaxaca_mean_distance}
\end{table}

\begin{figure}[tbp]
\centering
\includegraphics[width=0.9\textwidth]{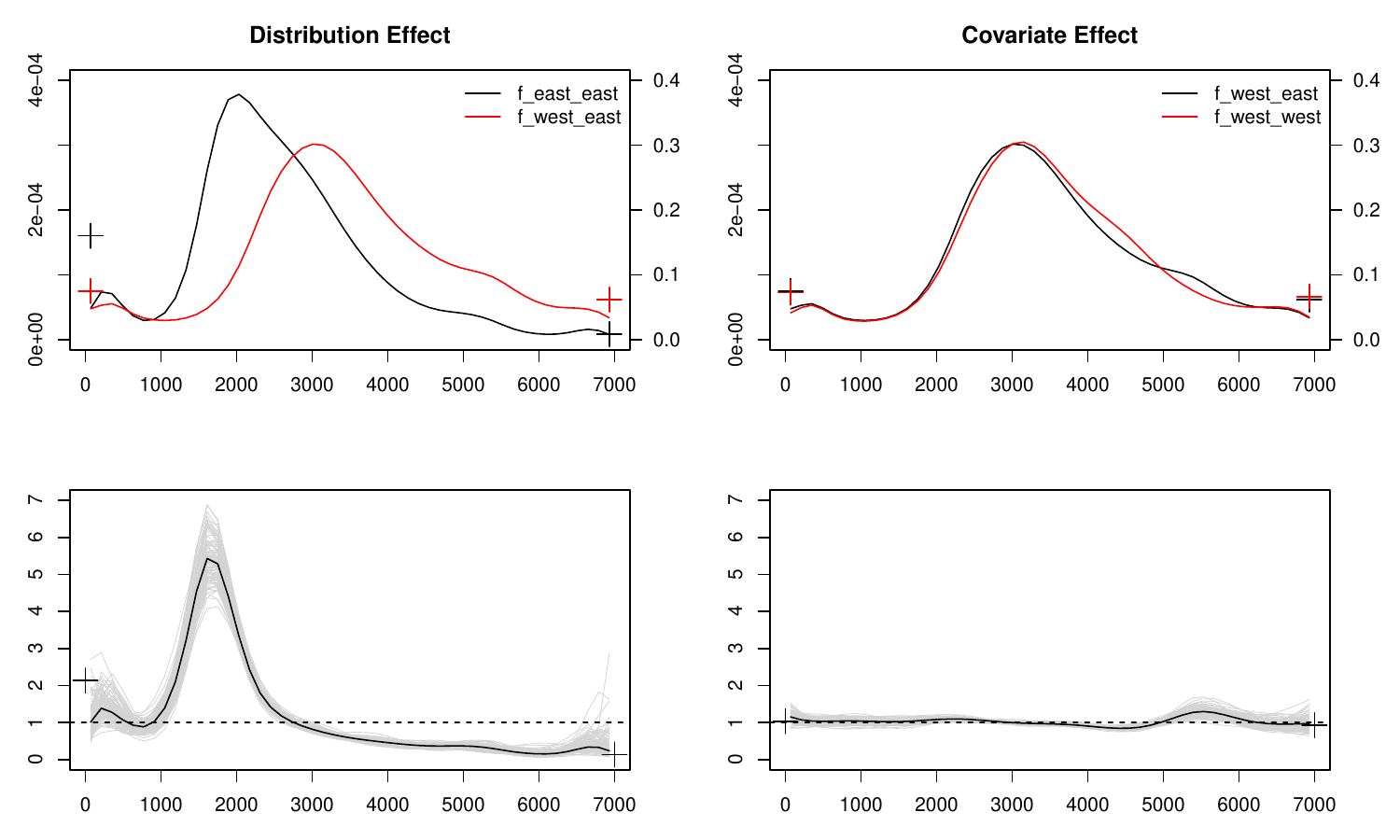}
\caption{Decomposition of the total density effect into distribution effect (left panels) and covariate effect (right panels) for men in the year 2001. The top panels show the estimated counterfactual densities, $f_{Y_{\langle\text{East},\text{East}\rangle}}$, $f_{Y_{\langle\text{West},\text{East}\rangle}}$, and $f_{Y_{\langle\text{West},\text{West}\rangle}}$. The lower panels show the estimated density effects $\operatorname{DE}(y)$ and $\operatorname{CE}(y)$, with 100 draws from the 95\% confidence region.}
\label{fig:male01}
\end{figure}

\begin{figure}[tbp]
\centering
\includegraphics[width=0.9\textwidth]{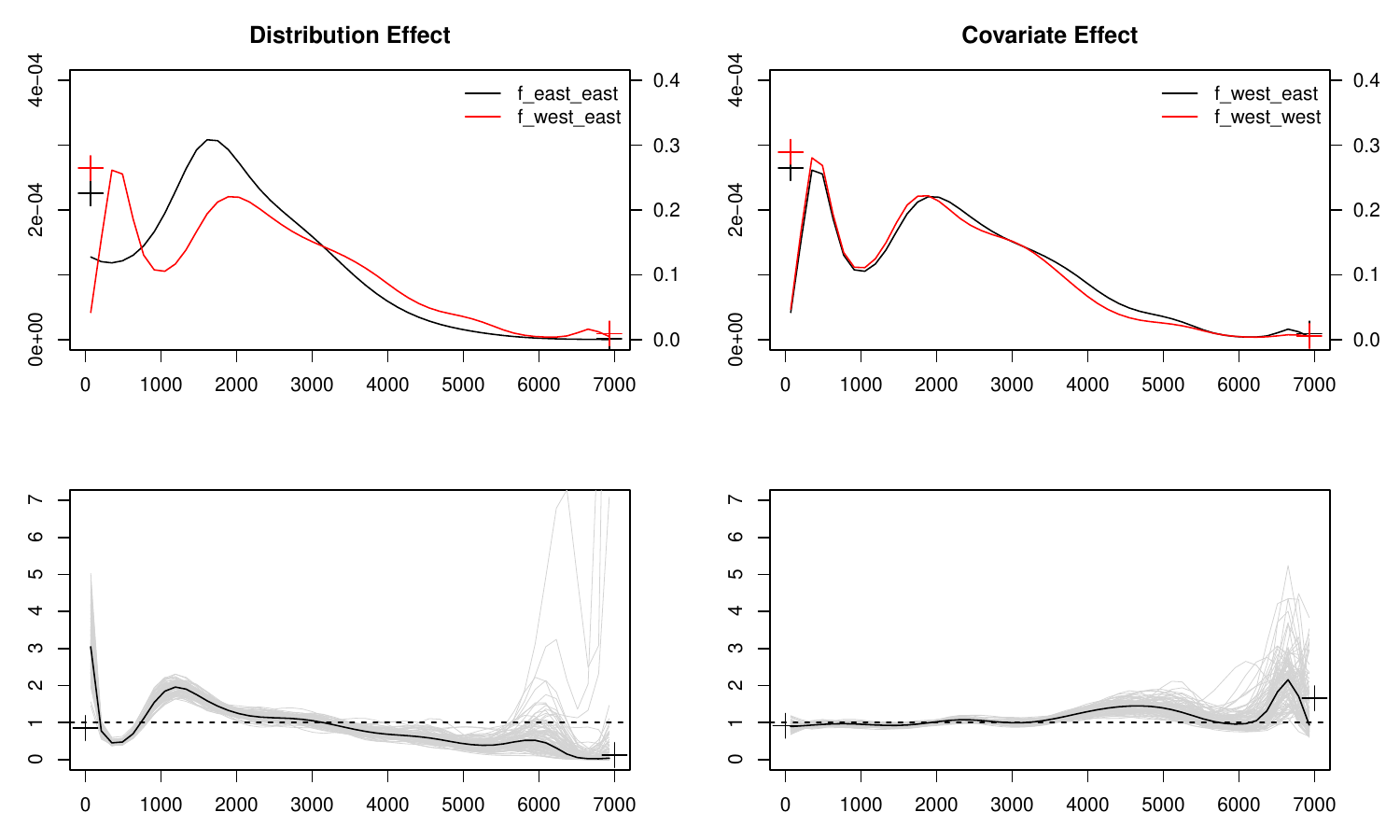}
\caption{Decomposition of the total density effect into distribution effect (left panels) and covariate effect (right panels) for women in the year 2001. The top panels show the estimated counterfactual densities, $f_{Y_{\langle\text{East},\text{East}\rangle}}$, $f_{Y_{\langle\text{West},\text{East}\rangle}}$, and $f_{Y_{\langle\text{West},\text{West}\rangle}}$. The lower panels show the estimated density effects $\operatorname{DE}(y)$ and $\operatorname{CE}(y)$, with 100 draws from the 95\% confidence region.}
\label{fig:female01}
\end{figure}

In the previous analysis, we assumed that the sex variable enters additively in the two income regressions. However, the decomposition of the East--West income gap may be fundamentally different for men and women. We therefore present additional estimation results for counterfactual densities and density effects based on separate estimations for men and women for 2001. Figures \ref{fig:male01} and \ref{fig:female01} indeed show that the East--West gap is a more pronounced issue for the male population. This can partly be explained by the larger share of part-time work for women in the West compared to the East. Indeed, one can see that the distribution effect is below one in the lower regions of the income distribution. I.e., relatively more women have a low income in West Germany than in East Germany. In contrast, we see similar but less strong effects for women in the upper tail regions of the income distribution.

In the following, we want to further analyze the covariate effect by isolating the impact of single variables using equation (\ref{eq:CEj}). We refer to Figure~\ref{fig:covariate01} in the supplementary material for the contributions to the covariate effect in 2001 of the variables education, $\operatorname{CE}_{\text{edu}}(y)$, and age, $\operatorname{CE}_{\text{age}}(y)$. For education, the contribution to the covariate effect is significantly below one in the lower income regions, which implies that the share of East Germans in these regions is higher despite and not because of differences in education. For age, we do not observe any significant effects in any direction.

Finally, in Section~\ref{subsec:application_industry} of the supplement, we include a robustness check by including an additional categorical variable controlling for the industrial sector of the main job. The inclusion leads to an endogeneity issue, since having an industry code presupposes the employment of the person, which is why we do not include it in our main empirical analysis. Due to this issue, we need to restrict the analysis to the continuous part of the income distribution. We show that the results are indeed robust towards controlling for industry, i.e., the inclusion of the variable cannot explain the remaining discrepancies in the income distributions between East and West.

\section{Conclusion}\label{sec:conclusion}

In this paper, we presented a new framework for conducting causal inference based on counterfactual densities. The approach is based on a multiplicative Oaxaca--Blinder type decomposition of the densities of the treatment and control groups into a distribution and covariate effect. To estimate the conditional densities, we rely on the Bayes Hilbert space additive density regression model of \cite{maier2024conditional}. As an application of our approach, we analyze the German East--West income gap. We find that differences between income densities decreased over time and that the decline can mainly be attributed to changes in the conditional distribution. In contrast, differences in the covariate distribution between East and West only play a minor role. Additionally, we find that the East--West gap is much more pronounced for the male sub-population.

A major advantage of our density-based approach is that it is able to capture high degrees of heterogeneity in the causal effects. Further, visualization of the counterfactual densities and corresponding effects allows for an intuitive interpretation of the modeled effects. Compared to alternative approaches based on quantiles, our approach allows for mixed types of the dependent variable. There are some limitations to our approach, which can be the subject of future research. First, the computational complexity of the Poisson estimation problem can be substantial in the case of large sample sizes and continuous explanatory variables. Second, it is currently assumed that the underlying additive density regression model is correctly specified. Future research could analyze the effect of misspecification on the estimation results.

\section*{Conflicts of Interest}
The authors declare no conflict of interests.

\section*{Acknowledgments}
Funded by the Deutsche Forschungsgemeinschaft (DFG, German Research Foundation) - Project number 513634041.

{
%\small
\onehalfspacing
\bibliography{literature.bib}
\bibliographystyle{apalike}
}

\newpage
\appendix

\begin{center}
    {\Huge \textbf{Supplementary Material}}
\end{center}

\section{Additional Estimation Results}\label{sec:application_other}

\subsection{Additional Figures}

This section contains additional Figures of the estimated counterfactual densities and density effects.

\begin{figure}[ht]
\centering
\includegraphics[width=0.8\textwidth]{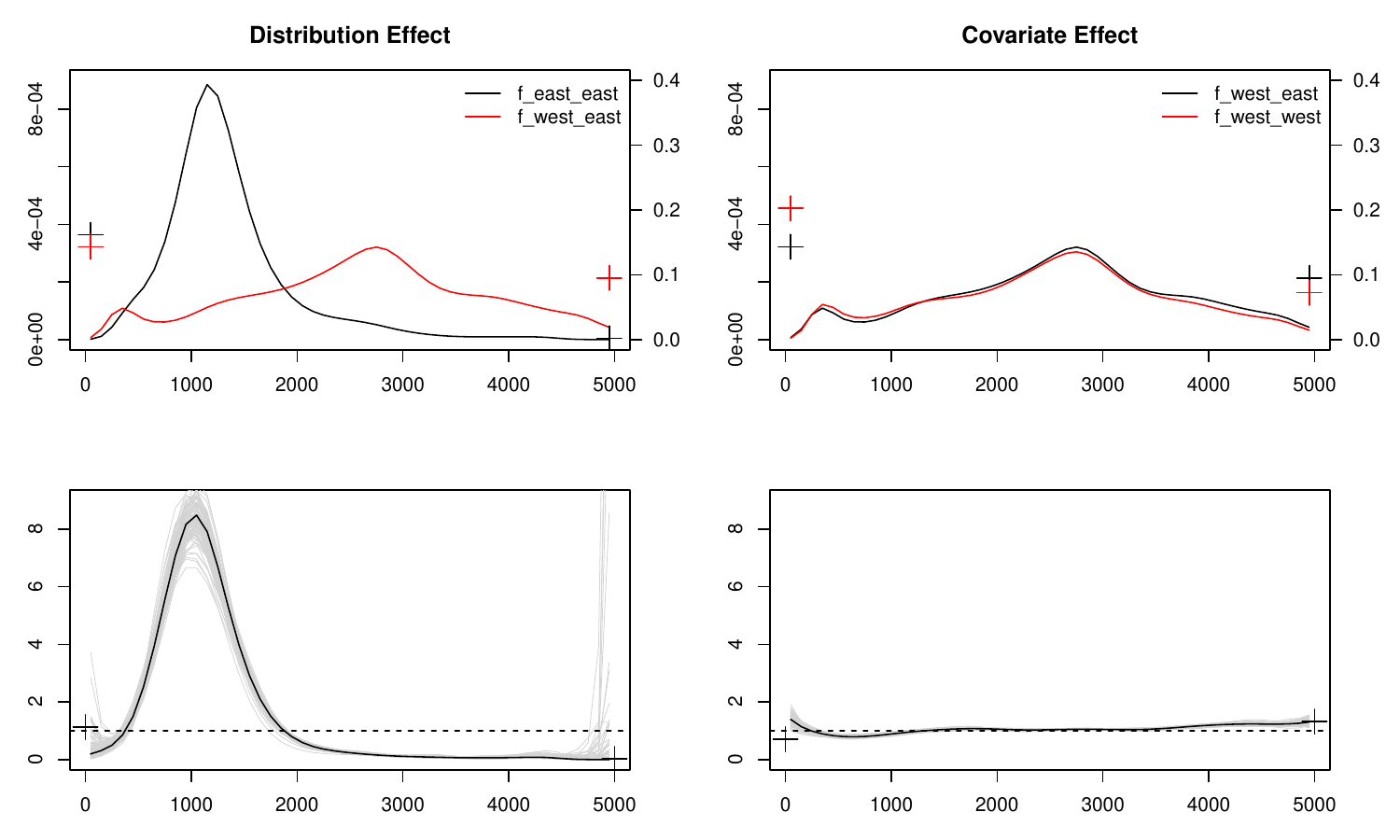}
\caption{Decomposition of the total density effect into distribution effect (left panels) and covariate effect (right panels) for the year 1991. The top panels show the estimated counterfactual densities, $f_{Y_{\langle\text{East},\text{East}\rangle}}$, $f_{Y_{\langle\text{West},\text{East}\rangle}}$, and $f_{Y_{\langle\text{West},\text{West}\rangle}}$. The lower panels show the estimated density effects $\operatorname{DE}(y)$ and $\operatorname{CE}(y)$ with 100 draws from the 95\% confidence region.}
\label{fig:counter91}
\end{figure}

\begin{figure}[ht]
\centering
\includegraphics[width=0.8\textwidth]{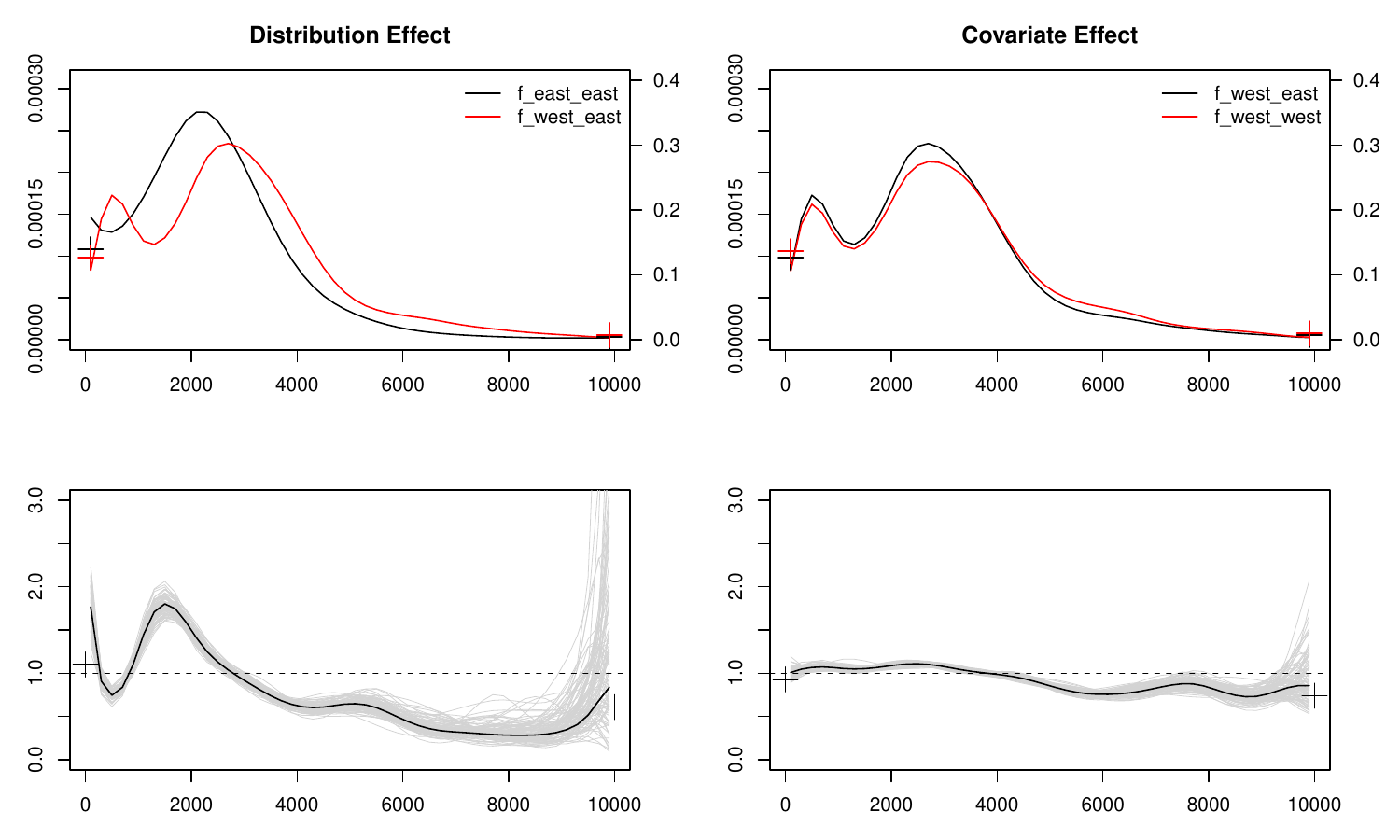}
\caption{Decomposition of the total density effect into distribution effect (left panels) and covariate effect (right panels) for the year 2011. The top panels show the estimated counterfactual densities, $f_{Y_{\langle\text{East},\text{East}\rangle}}$, $f_{Y_{\langle\text{West},\text{East}\rangle}}$, and $f_{Y_{\langle\text{West},\text{West}\rangle}}$. The lower panels show the estimated density effects $\operatorname{DE}(y)$ and $\operatorname{CE}(y)$ with 100 draws from the 95\% confidence region.}
\label{fig:counter11}
\end{figure}

\begin{figure}[ht]
\centering
\includegraphics[width=0.8\textwidth]{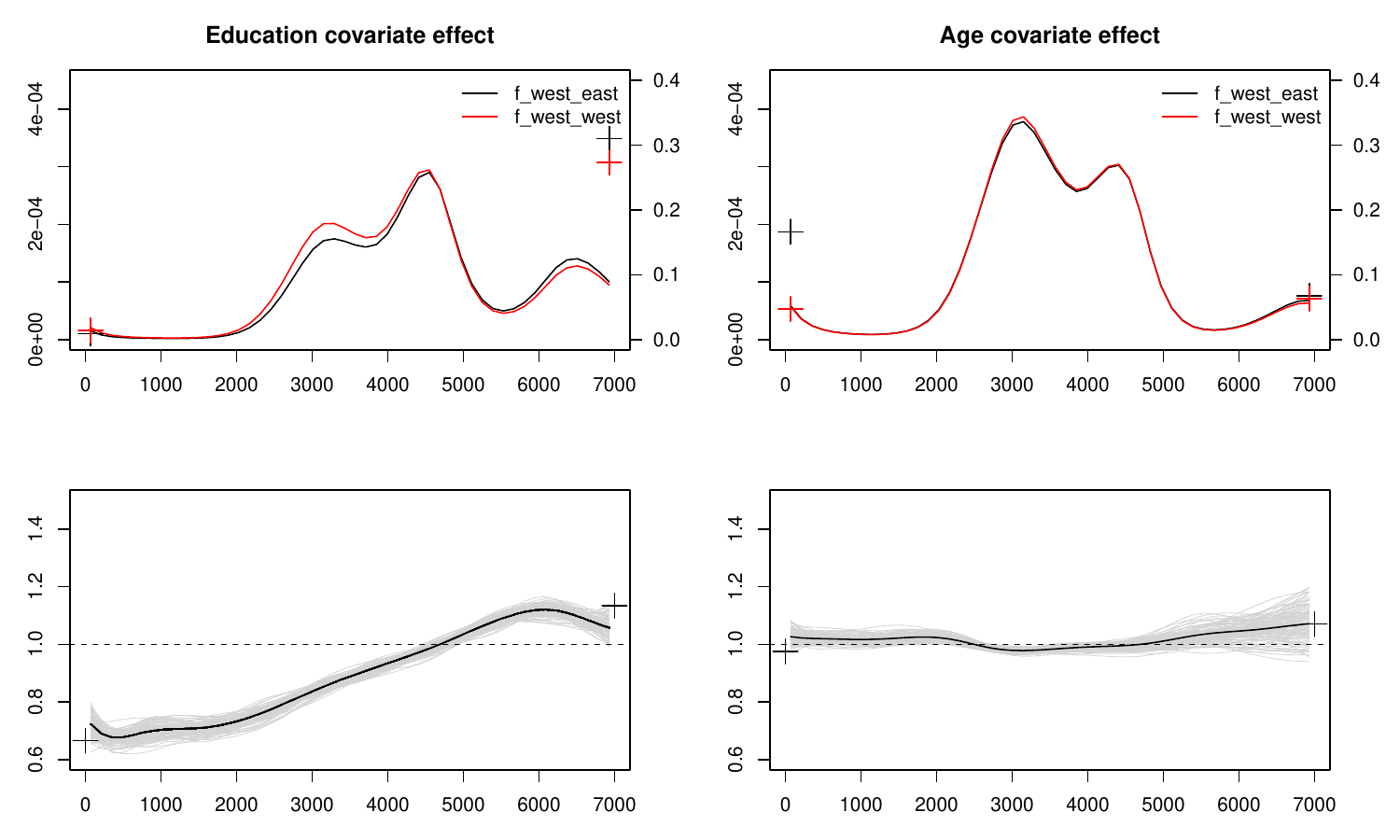}
\caption{Contribution to the Covariate effect of variables education (left panel) and age (right panel) in the year 2001. The top panels show the estimated counterfactual densities and the lower panels show the estimated covariate effects $\operatorname{CE}_{edu}(y)$ and $\operatorname{CE}_{age}(y)$, with 100 draws from the 95\% confidence region.}
\label{fig:covariate01}
\end{figure}

\FloatBarrier
\subsection{Robustness Check: Impact of Industry}\label{subsec:application_industry}

In this subsection, we provide a robustness check for our results by including an additional industry variable. This variable introduces an endogeneity problem, since membership in a certain industry presupposes that the person is employed and has a positive income. Nonetheless, information on industry may help explain the observed differences in income densities between East and West. For this purpose, we restrict our analysis to the positive part of income, discarding all individuals with zero income. For the industry categorization, we use the first digit of the variables `pgkldb92' for 1991, and `pgkldb10' for 2021.

\begin{figure}[ht]
\centering
\includegraphics[width=0.8\textwidth]{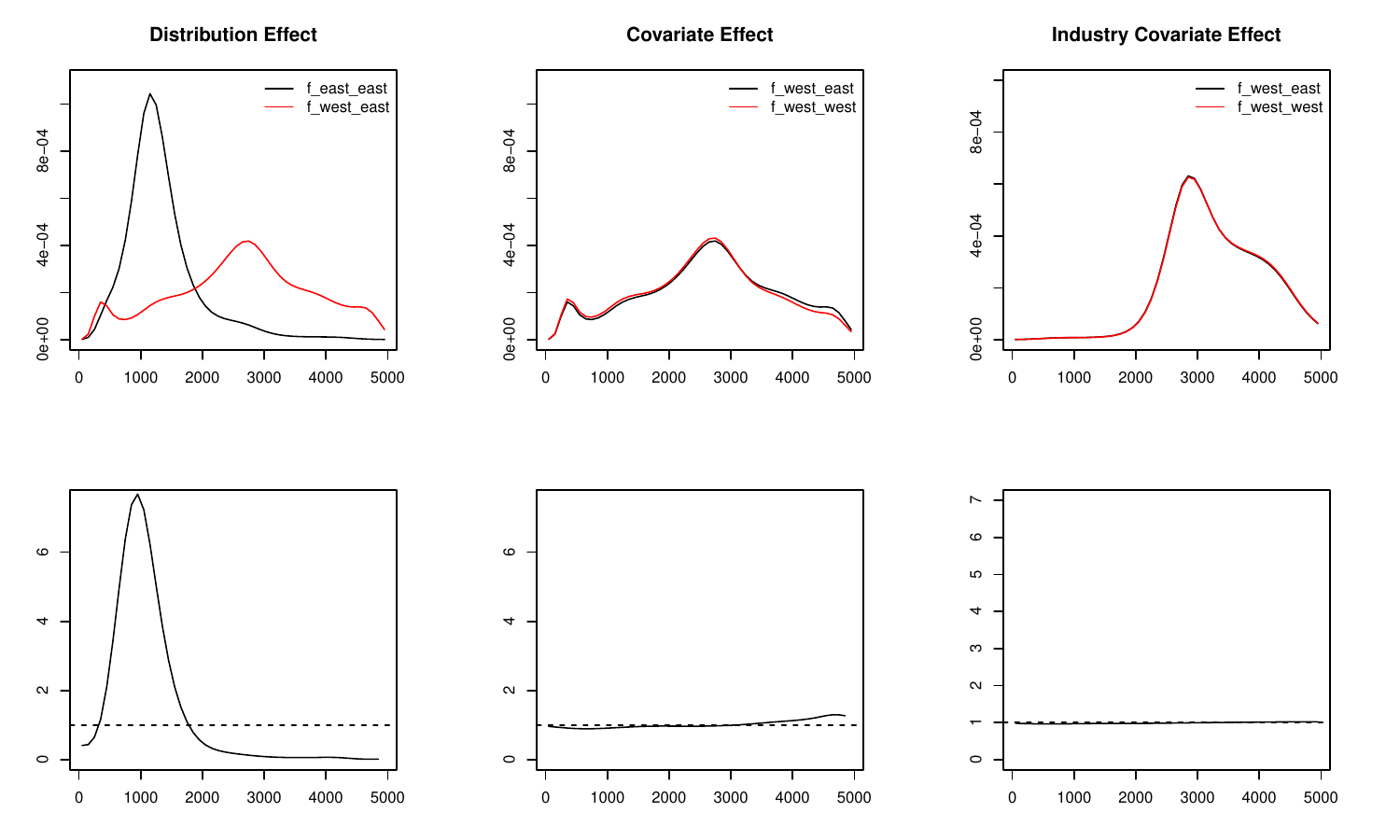}
\caption{Distribution effect, covariate effect and industry's contribution to the covariate effect for the estimation of the continuous part of the income densities in 1991.}
\label{fig:continuous91}
\end{figure}

\begin{figure}[ht]
\centering
\includegraphics[width=0.8\textwidth]{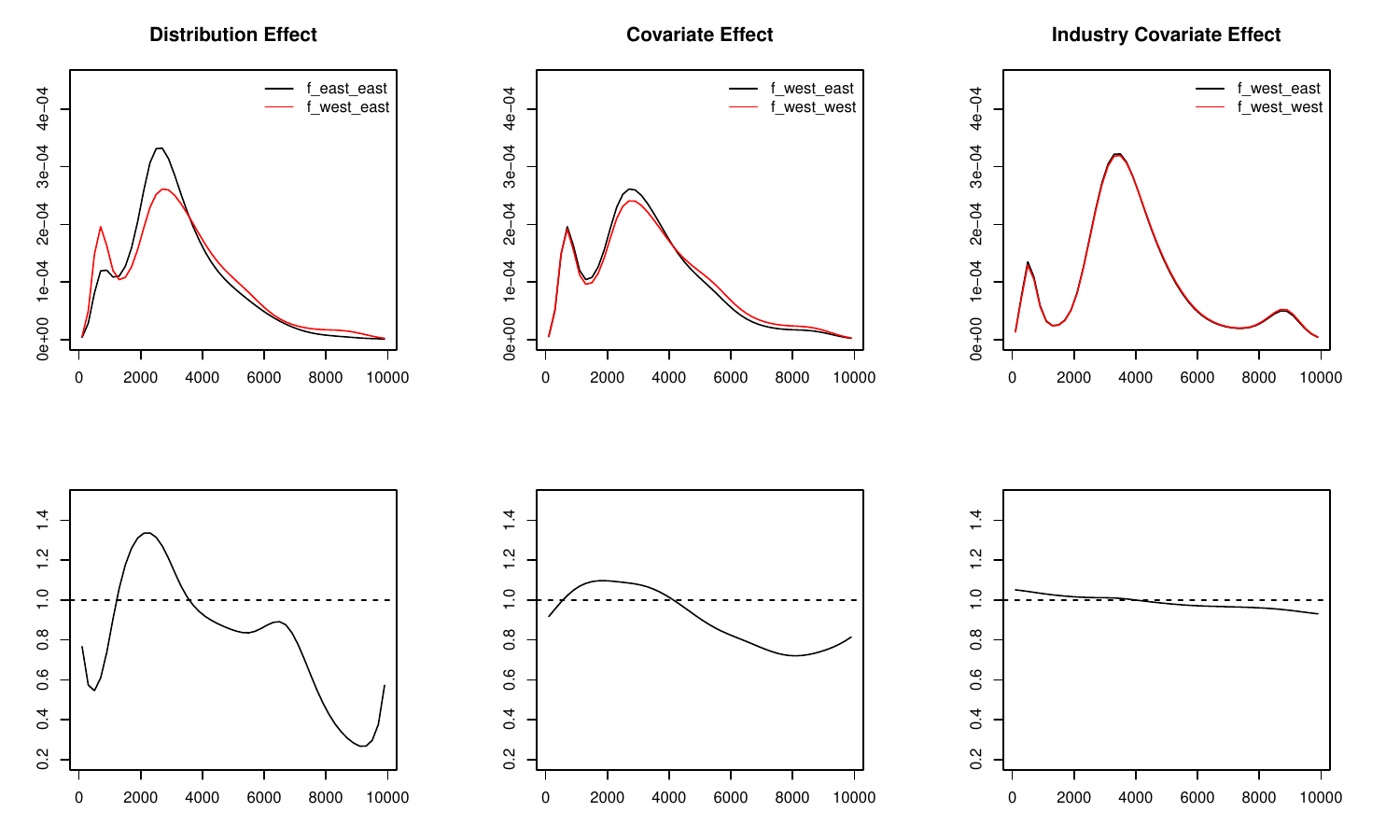}
\caption{Distribution effect, covariate effect and industry's contribution to the covariate effect for the estimation of the continuous part of the income densities in 2021.}
\label{fig:continuous21}
\end{figure}

See Figures \ref{fig:continuous91} and \ref{fig:continuous21} for the estimation results for the years 1991 and 2021, respectively. We observe that the observed discrepancy in (positive) income densities is again dominated by the distribution effect. The inclusion of the industry variable did not fundamentally change the results. In particular, when looking at the variable's contribution to the covariate effect, we conclude that the role is rather minor.

\section{Additional Simulation Details}\label{sec:appendix_sim}

This section provides additional details about the simulation settings and estimation methods used in Section~\ref{sec:simulation} in the main text.

For both the treatment and control group, we simulate $n$ observations for the covariates from a multinomial distribution. For the former, we consider uniform class probabilities. For the latter, we set the class probabilities to $(0.25, 0.2, 0.14, 0.125, 0.095, 0.008, 0.06, 0.05)$. I.e., we have $8$ combinations of covariates, with the first category having binary covariates $x_1=x_2=x_3=1$, the second category having covariates $x_1=x_2=1$, $x_3=2$, etc. For each simulated covariate observation, we simulate the corresponding dependent variable from the conditional density, which we choose to be a beta distribution with parameters $\alpha_1=\beta_1=(1,5,5,9,2,6,6,10)$ for the treatment group, and $\alpha_0=(1,10,2,11,2,11,3,12)$ and $\beta_0=(1,2,10,11,2,3,11, 12)$ for the control group. The true counterfactual densities are therefore mixtures of beta distributions.

For the estimation of the conditional densities using the Bayes Hilbert space approach, we first discretize the support of the dependent variable into 50 equally sized histogram bins. The number of spline basis function is fixed at 12. The estimation is carried out without imposing additional penalty terms on the estimated coefficients. The benchmark approach involving kernel density estimation uses a Gaussian kernel and the bandwidth is selected according to Silverman's rule of thumb.

\end{document}